\documentclass[11pt,a4paper]{article}
\pdfoutput=1

\usepackage{jcappub}
\usepackage{graphicx}
\usepackage{epsfig}
\usepackage{subfigure}
\usepackage{color}
\usepackage{comment}
\usepackage{slashed}
\usepackage{soul}
\usepackage{tabularx}
\usepackage{physics}

\usepackage[autostyle]{csquotes}

\newcommand{\be}{\begin{equation}} 
\newcommand{\ee}{\end{equation}}
\newcommand{\bea}{\begin{equation}\begin{aligned}} 
\newcommand{\eea}{\end{aligned}\end{equation}}
\newcommand{\ber}{\begin{eqnarray}}
\newcommand{\ear}{\end{eqnarray}}

\def\lsim{\mathrel{\raise.3ex\hbox{$<$\kern-.75em\lower1ex\hbox{$\sim$}}}}
\def\gsim{\mathrel{\raise.3ex\hbox{$>$\kern-.75em\lower1ex\hbox{$\sim$}}}}

\usepackage{array}
\newcolumntype{C}[1]{>{\centering\let\newline\\\arraybackslash\hspace{0pt}}m{#1}}

\newcommand{\td}{{\rm d}}

\newcommand{\mpl}{M_{\rm P}}

\newcommand{\eps}{\epsilon}

\newcommand{\bigO}{\mathcal{O}}

\begin{document}

\title{Tachyonic Preheating in \\ Palatini $\bf{R^2}$ Inflation}

\author{Alexandros Karam,}
\author{Eemeli Tomberg}
\author{and Hardi Veerm\"{a}e}
\affiliation{Laboratory of High Energy and Computational Physics, 
National Institute of Chemical Physics and Biophysics, R{\"a}vala pst.~10, Tallinn, 10143, Estonia}

\emailAdd{alexandros.karam@kbfi.ee}
\emailAdd{eemeli.tomberg@kbfi.ee}
\emailAdd{hardi.veermae@cern.ch}

\abstract{
We study preheating in the Palatini formalism with a quadratic inflaton potential and an added $\alpha R^2$ term. In such models, the oscillating inflaton field repeatedly returns to the plateau of the Einstein frame potential, on which the tachyonic instability fragments the inflaton condensate within less than an e-fold. We find that tachyonic preheating takes place when $\alpha \gtrsim 10^{13}$ and that the energy density of the fragmented field grows with the rate $\Gamma/H \approx 0.011 \times \alpha^{0.31}$. The model extends the family of plateau models with similar preheating behaviour. Although it contains non-canonical quartic kinetic terms in the Einstein frame, we show that, in the first approximation, these can be neglected during both preheating and inflation.
}

\maketitle

\section{Introduction}

The theory of cosmic inflation~\cite{Starobinsky:1980te, Guth:1980zm, Linde:1981mu, Albrecht:1982wi, Linde:1983gd, Lyth:1998xn} is an attractive paradigm for the very early Universe since it provides a solution to some outstanding puzzles of the standard hot Big Bang cosmology such as the flatness and horizon problems. Most importantly, during this epoch of exponential expansion, quantum fluctuations seed primordial inhomogeneities which will grow into the large scale structure we observe today.

The latest data from the Planck satellite~\cite{Akrami:2018odb} have severely constrained the allowed values for the scalar spectral index $n_s$ and the tensor-to-scalar ratio $r$. As a consequence, simple inflationary models with monomial potentials have been ruled out. However, adding an $R^2$ term and studying these models in the Palatini formalism can significantly lower the value of $r$~\cite{Enckell:2018hmo, Antoniadis:2018ywb} (see also~\cite{ Antoniadis:2018yfq, Tenkanen:2019jiq, Tenkanen:2019wsd, Gialamas:2019nly, Tenkanen:2020dge, Tenkanen:2020cvw, Lloyd-Stubbs:2020pvx, Antoniadis:2020dfq, Das:2020kff, Gialamas:2020snr}) and still lead to single-field inflation, unlike in the metric formalism in which adding an $R^2$ term would introduce an additional scalar degree of freedom.

In the Palatini approach~\cite{Palatini1919, Ferraris1982}, the spacetime connection is not the usual Levi-Civita one as in the metric formalism, but rather it is assumed to be an independent variable. Therefore, the action has to be varied with respect to both the metric and the connection. Although the two formulations are equivalent for general relativity, differences arise when the inflaton is non-minimally coupled to gravity~\cite{Bauer:2008zj, Bauer:2010bu, Tamanini:2010uq, Bauer:2010jg, Rasanen:2017ivk, Tenkanen:2017jih, Racioppi:2017spw, Markkanen:2017tun, Jarv:2017azx, Fu:2017iqg, Racioppi:2018zoy, Carrilho:2018ffi, Kozak:2018vlp, Rasanen:2018fom, Rasanen:2018ihz, Almeida:2018oid, Shimada:2018lnm, Takahashi:2018brt, Jinno:2018jei, Rubio:2019ypq, Bostan:2019uvv, Bostan:2019wsd, Tenkanen:2019xzn, Racioppi:2019jsp, Tenkanen:2020dge, Shaposhnikov:2020fdv, Borowiec:2020lfx, Jarv:2020qqm, Karam:2020rpa, McDonald:2020lpz, Langvik:2020nrs, Shaposhnikov:2020gts, Shaposhnikov:2020frq, Gialamas:2020vto, Verner:2020gfa, Enckell:2020lvn, Reyimuaji:2020goi} or in $f(R)$ theories of gravity~\cite{Olmo:2011uz, Bombacigno:2018tyw, Enckell:2018hmo, Antoniadis:2018ywb, Antoniadis:2018yfq, Tenkanen:2019jiq, Edery:2019txq, Giovannini:2019mgk, Tenkanen:2019wsd, Gialamas:2019nly, Tenkanen:2020dge, Tenkanen:2020cvw, Lloyd-Stubbs:2020pvx, Antoniadis:2020dfq, Ghilencea:2020piz, Das:2020kff, Gialamas:2020snr, Ghilencea:2020rxc, Bekov:2020dww, Dimopoulos:2020pas, Gomez:2020rnq}. For example, while in the metric formalism non-minimally coupled Higgs inflation (omitting quantum corrections) predicts a value of the non-minimal coupling $\xi \sim\mathcal{O}(10^4)$, the Palatini version of the model predicts $\xi \sim \mathcal{O}(10^9)$. Similarly, as mentioned above, the effect of an $\alpha R^2$ term in the Palatini formalism is to flatten the inflationary potential and reduce the scale of inflation (compared to the same model studied in the metric formalism without the $\alpha R^2$ term), thus modifying the prediction for the tensor-to-scalar ratio $r$.

A typical feature of inflationary models in the Palatini approach is that the Einstein frame potential becomes a plateau for large field values. As a result, it is possible to push the Hubble scale below the frequency of inflaton oscillations so that, after inflation ends, the oscillating field can repeatedly return to the plateau. Because the field perturbations possess a strong tachyonic instability at the edge of the plateau, efficient production of inflaton particles can take place and dominate preheating~\cite{Rubio:2019ypq, Karam:2020rpa}. Due to this flattening effect, Palatini $R^2$ models provide a natural framework for tachyonic preheating.

In this paper, we consider a minimal model in which the inflaton is governed by a quadratic potential and extend it with an $R^2$ term in the Palatini formalism. While the quadratic model usually predicts $r=0.13$, which is excluded by Planck~\cite{Akrami:2018odb}, the effect of the $\alpha R^2$ term is to flatten the potential and lower the value of $r$ as $\alpha$ increases, rendering the model viable again~\cite{Enckell:2018hmo, Antoniadis:2018yfq, Tenkanen:2019wsd, Gialamas:2020snr}. At the same time, a higher-order quartic kinetic term is generated in the Einstein frame. This term can be safely ignored during slow-roll inflation~\cite{Enckell:2018hmo, Tenkanen:2020cvw}, and we show that it is sub-dominant also during preheating. This is one of the main results of our work. 
We study the preheating process numerically in the linear approximation, complemented by analytical results, and compute the preheating timescale as a function of $\alpha$, as well as the spectrum of perturbations as a function of the comoving wavenumber $k$.

The paper is organized as follows. In section \ref{sec:models}, we review the general properties of Palatini $R^2$ models and introduce the specific model considered in this work. In section \ref{sec:background_evolution}, we review the inflaton field's behaviour during inflation and discuss its evolution in the subsequent preheating stage. In section \ref{sec:particle_production}, we study particle production during preheating and present our main results. We also comment on the UV cut-off scale of the model and interactions with Standard Model (SM) particles. Finally, we conclude in section \ref{sec:conclusions}. Some technical details are gathered in the appendices. We use natural units $\hbar = c = k_\mathrm{B} = \mpl^2 = 1$ and the metric signature $(-,\!+,\!+,\!+)$.

\section{Palatini $R^2$ models}
\label{sec:models}

Let us start by considering a scalar field $\varphi$, the metric $g_{\mu\nu}$, and a symmetric connection $\Gamma^{\rho}_{\mu\nu}$, treated as independent variables in the Jordan frame action\footnote{We will refer to the frame in Eq.~\eqref{action1} as the Jordan frame even when $A = 0$.}
\be \label{action1}
  S = \int\dd^4 x \sqrt{-g} \left[ \frac{1}{2} R + \frac{\alpha}{2} R^2 + \frac{1}{2} A(\varphi) R - \frac{1}{2} (\partial \varphi)^2 - V(\varphi) \right] \ .
\ee
Here $g$ is the determinant of $g_{\mu\nu}$ and $R=g^{\mu\nu} R^{\rho}_{\ \, \mu\rho\nu}(\Gamma, \partial\Gamma)$ is the Ricci scalar (the Riemann tensor is built from the connection alone in the Palatini formulation). By introducing an auxiliary scalar $\chi$, we can rewrite the $R^2$ term as $\chi^2 R - \chi^4/2$. By performing a Weyl rescaling of the metric
\be\label{eq:conf}
    g_{\mu\nu} \rightarrow \Omega^{-2} g_{\mu\nu} = [ 1 + \alpha \chi^2 + A(\varphi) ]^{-1} g_{\mu\nu} 
\ee
and eliminating the auxiliary field $\chi$, the action becomes~\cite{Enckell:2018hmo, Antoniadis:2018ywb}
\be \label{action5}
    S   =   \int\dd^4 x \sqrt{-g} \left[ \frac{1}{2} R - \frac{1}{2} \frac{1 - 8\alpha U}{ 1+A } \left( \partial \varphi \right)^2 + \frac{\alpha}{2} \frac{1 - 8\alpha U}{ (1+A)^2 } \left( \partial \varphi \right)^4  - U \right] \ ,
\ee
where we defined
\be \label{Ub}
  U \equiv \frac{V}{(1+A)^2 + 8 \alpha V} \, .
\ee
A canonical (quadratic) kinetic term can then be obtained by redefining the field via
\be \label{fieldredef}
    \frac{\dd \varphi}{\dd \phi} = \sqrt{ \frac{1+A}{1 - 8\alpha U} } \ ,
\ee
yielding the action 
\be \label{action6}
    S = \int\dd^4 x \sqrt{-g} \left[ \frac{1}{2} R - \frac{1}{2} \left( \partial \phi \right)^2 + \frac{1}{2} \frac{\alpha}{1 - 8 \alpha U} \left( \partial \phi \right)^4  - U  \right] \ .
\ee
The $R^2$ term has been translated into a higher-order kinetic term and a modification of the Jordan frame potential. Because a negative $\alpha$ would lead to negative kinetic energy, implying an unstable system, we take $\alpha>0$. In this case, we see that, when the Jordan frame potential is positive, i.e., $V \geq 0$, then the Einstein frame potential is bounded as $0 < U < (8\alpha)^{-1}$. This property avoids singularities in the quartic kinetic term and leads to the flattening of any monotonously growing Jordan frame potential at large field values.

\begin{figure}[ht]
\centering
\includegraphics[width=.8\linewidth]{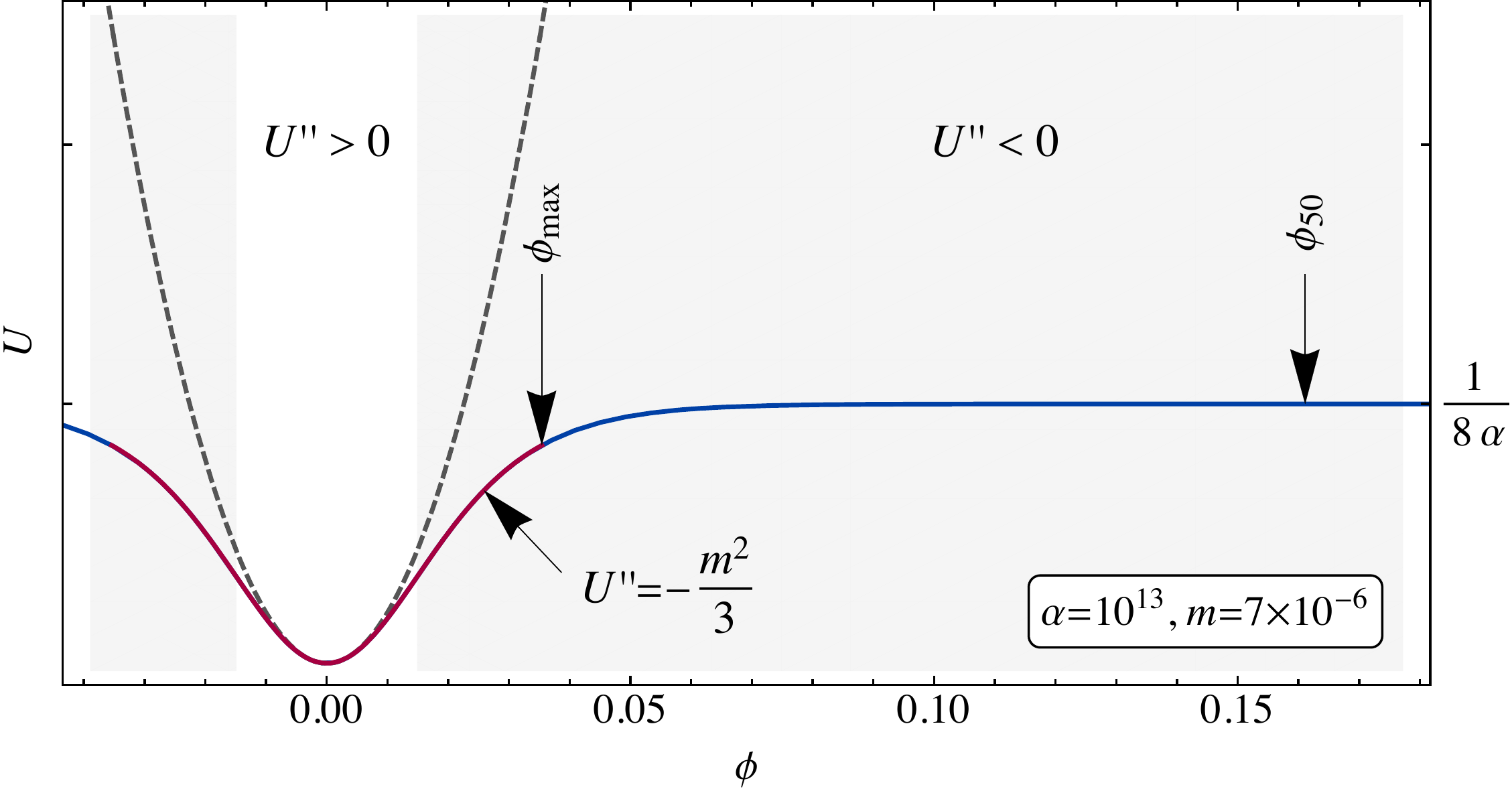}
\caption{The potential as a function of the value of the inflaton field. The Einstein frame potential \eqref{eq:pot_E} with $m = 7\times 10^{-6}$ and $\alpha = 10^{13}$ is shown by the solid line, while the dashed line depicts the Jordan frame potential $m^2\phi^2/2$. $\phi_{50}$ is the field value corresponding to 50 $e$-folds of inflation. The region in which the tachyonic instability is active (grey shaded area) overlaps with the region in which the inflaton will oscillate (solid red line). $\phi_{\max}$ shows the maximal amplitude of oscillations. The leftmost arrow shows the point at which the tachyonic instability is maximal, $U'' = -m^2/3$.}

\label{fig:1}
\end{figure}

We will keep the theoretical discussion as general as possible. However, for the sake of concreteness, in the numerical examples we will consider the case of a minimally coupled, free scalar field~\cite{Linde:1983gd, Linde:1984st, Madsen:1988xe}
\be \label{eq:pot_J}
    V(\varphi) = \frac{1}{2} m^2 \varphi^2 \,, \qquad A(\varphi) = 0 \,.
\ee
With a minimal coupling to gravity, this is the only stable potential which gives the correct CMB predictions and is renormalizable in the Jordan frame. This is because adding an $R^2$ term in the Palatini formulation does not change $n_s$~\cite{Enckell:2018hmo} and, therefore, a quartic potential would be in conflict with observations as it predicts a scalar spectral index $n_s \approx 0.94$, which is excluded~\cite{Akrami:2018odb}. The Einstein frame field, defined through~\eqref{fieldredef}, is
\be 
    \phi 
    = \phi_0 \sinh^{-1}(\varphi/\phi_0) \ , \qquad
    \phi_0 \equiv \frac{1}{2 m \sqrt{\alpha}},
\ee
where the scale $\phi_0$ roughly determines the onset of the plateau.
The Einstein frame potential~\eqref{Ub} thus reads
\be \label{eq:pot_E}
    U 
    = \frac{1}{8 \alpha}\tanh^2 (\phi/\phi_0) \ .
\ee
This potential is depicted in Fig.~\ref{fig:1} for $m = 7\times 10^{-6}$ and $\alpha = 10^{13}$. As discussed above, the effect of the $R^2$ term is to produce an Einstein frame potential (solid line) by flattening the Jordan frame potential (grey dashed line) to $(8\alpha)^{-1}$, while preserving the shape of the potential at low field values, i.e., when $U \ll (8\alpha)^{-1}$. Due to this flattening, the effective mass $U''$ of perturbations can acquire large negative values triggering a tachyonic instability. In our current example, $U''$ reaches its minimal value $-m^2/3$ at the point indicated in Fig.~\ref{fig:1}. Another peculiar consequence of this flattening effect is that, for large enough $\alpha$, the frequency of oscillations, currently $\mathcal{O}(m)$, can significantly exceed the Hubble rate $H \approx (24\alpha)^{-1/2}$. As shown in Fig. \ref{fig:1}, in this case, slow-roll inflation ends while the field is essentially still on the plateau. Thus, as the field starts to oscillate, it can repeatedly re-enter the tachyonic region resulting in a very effective preheating.

Compared to metric theories, no extra degrees of freedom appear in the Palatini approach. This is because in the Palatini approach no kinetic term arises for the auxiliary field $\chi$ and thus its equation of motion reduces to just a constraint which can be used to eliminate it altogether. Consequently, the Einstein frame action~\eqref{action6} contains only one field, in contrast to the metric case where the resulting action would contain two scalar fields.  Moreover, the $R^2$ term does not flatten the Einstein frame potential in metric theories. Thus, actions of the form \eqref{action1} do not generally support tachyonic preheating in the metric formalism unless the Jordan frame potential already has a shape required for tachyonic preheating. In particular, the Einstein frame potential \eqref{eq:pot_E} does not follow from the Jordan frame action \eqref{action1} with the Jordan frame potential \eqref{eq:pot_J}. Preheating through parametric resonance in metric $R^2$ gravity with the action \eqref{eq:pot_J} has been studied in~\cite{vandeBruck:2016leo} (see also~\cite{DeCross:2015uza, DeCross:2016fdz, Mori:2017caa, Pi:2017gih, Canko:2019mud}).

The Einstein frame potential \eqref{eq:pot_E} has been considered in several earlier works both in the context of Palatini gravity~\cite{Antoniadis:2018ywb, Tenkanen:2019wsd, Gialamas:2019nly, Gialamas:2020snr} and in the context of the metric formulation~\cite{Kallosh:2013hoa, Kallosh:2013yoa, Ling:2021zlj}, where it is usually referred to as the T-model, but the single-field tachyonic preheating stage we encounter has not been studied before in detail for this potential. As discussed above, compared to metric theories, the Einstein frame action in the Palatini $R^2$ case has a manifestly non-canonical kinetic term implying modifications to the speed of sound. However, as we will find, the quartic kinetic term does not significantly affect the phenomenological predictions.

\section{Background evolution}
\label{sec:background_evolution}

Before considering particle production, we first study the evolution of the homogeneous and isotropic background field during inflation and preheating. Initially, the field is far on the plateau, and the Universe undergoes a period of slow-roll inflation. After its end, the field begins to oscillate around the minimum of the potential. When $\alpha$ is sufficiently large, inflation can end while the field is still on the plateau and therefore the field may repeatedly return to the plateau in the oscillating regime, as we will see shortly. This means that the field will spend a significant fraction of time in the tachyonic instability region boosting the growth of fluctuations, as seen previously in \cite{Rubio:2019ypq, Karam:2020rpa}. We will outline the classical evolution of the background without accounting for the feedback from these fluctuations.

In the spatially flat Friedmann--Robertson--Walker (FRW) Universe, the expansion is controlled by the Friedmann equation $3 H^2 = \rho$. The energy density and pressure of the field, derived from the action \eqref{action6}, are
\be\label{eq:rho_P}
    \rho = \frac{1}{2} \left( 1 + \frac{3}{2}B  \right) \dot\phi^2 + U, \qquad
    P = \frac{1}{2} \left( 1 + \frac{1}{2}B  \right) \dot\phi^2 - U \ ,
\ee
and the field equation reads
\be \label{FRW}
    \ddot\phi + 3 H c_s^2 \dot\phi + \frac{1 + 3B^2}{1 + 3B} U' = 0 \ , 
\ee
where we defined
\be \label{eq:B_cs}
    B \equiv \frac{2 \alpha \dot\phi^2}{1 - 8\alpha U},    
    \qquad
    c_s^2 = \frac{1+B}{1+3B} \ .
\ee
The parameter $B$ quantifies the relative contribution from the quartic kinetic term: when $B \ll 1$ ($B \gg 1$), the quartic kinetic term is negligible (dominant). The quantity $c_s$ corresponds to the speed of sound of perturbations~\cite{Garriga:1999vw}.

\subsection{Inflation}

Slow-roll inflation is characterized by the slow-roll parameters
\begin{equation}
    \epsilon_U \equiv \frac{1}{2}\left(\frac{U'}{U}\right)^2 \ , \qquad \eta_U \equiv \frac{U''}{U} \ ,
\end{equation}
which are small on the flat inflationary plateau, and the slow-roll approximation is valid. The slow-roll approximation gives the number of e-folds of expansion from field value $\phi$ to the end of inflation,
\be\label{eq:N_SR}
    N 
    \approx \int^{\phi}_{0} \frac{\dd \phi}{\sqrt{2 \epsilon_U}}
    \approx \frac{\phi_0^2}{4}\sinh^2{\frac{\phi}{\phi_0}} 
    \qquad \Rightarrow \qquad 
    \phi 
    \approx \frac{\phi_0}{2} \ln \frac{16N}{\phi_0^2} \, ,
\ee
where we took the limit $\phi_0 \ll 1$. This limit is needed for successful preheating, as we will see in section \ref{sec:particle_production}. 
Regarding the inflationary observables, as was shown for a general case in~\cite{Enckell:2018hmo}, the amplitude of the scalar power spectrum $A_s$ and the scalar spectral index $n_s$ remain unaffected by the $\alpha R^2$ term up to first order in the slow-roll parameters:
\be 
    n_s = 1 - 6 \epsilon_U + 2 \eta_U
    \approx 1 - \frac{2}{N_*}
    \ , \qquad
    A_s = \frac{U}{24\pi^2 \epsilon_U}
    \approx \frac{m^2 N^2_*}{6 \pi^2} \, .
\ee
Here $N_* \approx 50$ at the CMB pivot scale, so that the spectral index $n_s \approx 0.96$ is compatible with the Planck results~\cite{Akrami:2018odb}. The measured power spectrum strength $A_s = 2.1 \times 10^{-9}$ yields 
\be \label{eq:m_value}
    m = 7\times 10^{-6}
\ee
and we are left with a single free parameter $\alpha$ (or equivalently $\phi_0$). The $\alpha R^2$ term affects the expression for the tensor-to-scalar ratio $r$, which becomes
\be\label{eq:r}
    r 
    = 16\epsilon_U 
    \approx \frac{8}{N_* + 4 N^2_*/\phi_0^{2}} 
    \approx \frac{2\phi_0^2}{N^2_*} \,,
\ee
where the last approximation holds when $4N_* \gg \phi_0^{2}$ or, equivalently, $\alpha \gg 10^{-3} m^{-2} \approx 2\times 10^{7}$ as the mass is fixed. For smaller values of $\alpha$, the effect becomes negligible and we recover $r = 8/N_* \approx 0.16$, i.e. the prediction of standard quadratic inflation which is excluded. The tensor-to-scalar ratio $r$ is compatible with the Planck constraint $r<0.056$ when $\alpha \gtrsim 5 \times 10^7$, thus inflation must take place in the regime where the $\alpha R^2$ term is relevant.

Above, we neglected the quartic kinetic term in \eqref{action6} and set $B=0$---it was suggested in \cite{Enckell:2018hmo} that this is a good approximation during slow-roll. For the potential \eqref{eq:pot_J} we find that, in the slow-roll approximation and in the $\phi_0 \ll 1$, $N \gg 1$ limit, (see appendix \ref{sec:quartic_kinetic_term} for details)
\be
    B \approx \frac{1}{12N} \ll 1.
\ee
The effect of the quartic kinetic term on the inflationary observables is thus indeed small. However, it may still affect preheating dynamics after inflation. We will return to this point in section~\ref{sec:particle_production}.

\subsection{Post-inflationary oscillation}
\label{sec:oscillation}

After inflation, the inflaton starts to oscillate around the minimum of the potential. We will solve this phase numerically in section \ref{sec:particle_production}, but let us first gain some analytical understanding of the oscillations.

As the numerics show, in the limit of small $\phi_0$ (large $\alpha$), the frequency of oscillations $\omega$ is much bigger than the Hubble rate, $H \ll \omega$. The expansion of space can be neglected during a single oscillation, and we can work in the Minkowski limit $H \to 0$. The effect of the expansion on the energy density can be studied by considering temporal averages of the field equations. These can be recast in terms of the continuity equation
\be\label{eq:cont}
    \dot\rho + 3H (\rho + \bar P) = 0\,,
\ee
where barred quantities denote time averages. 

Before continuing, let us estimate the lower bound on $\alpha$ for the condition $H \ll \omega$ to be satisfied after inflation ends. In the model determined by Eq.~\eqref{eq:pot_J}, $\omega = \bigO(m)$ is a reasonable guess. Then, using the fact that on the plateau $H\approx (24 \alpha)^{-1/2}$, we obtain $\alpha \gg 10^{9}$. In the following, we will show that $\alpha \gtrsim 10^{13}$ to satisfy the requirements for tachyonic preheating.

The evolution of an homogeneous field on a flat background can be described as a mechanical system with the action $\int \td t P(\phi,\dot\phi)$ and a corresponding conserved energy $\rho(\phi,\dot\phi) = \dot\phi\partial P/\partial \dot\phi - P$,
which, for the model under consideration, are given by Eq.~\eqref{eq:rho_P}. 
Due to energy conservation, we can express the derivative of the field as $\dot\phi(\phi,\rho)$, where $\rho$ is now a constant of integration. Integrating the field equation gives the half-period of oscillations, and the time average of the pressure as
\be
    T_{1/2}(\rho) \equiv \int \td t = \int^{\phi_2}_{\phi_1} \frac{\td \phi}{\dot\phi(\phi,\rho)} \, , \qquad
    \bar P(\rho) \equiv \frac{1}{T_{1/2}(\rho)}\int^{\phi_2}_{\phi_1} \td t P(\phi,\dot\phi(\phi,\rho))\,,
\ee
where the turning points $\phi_{i}(\rho)$ are defined as the field values where the velocity vanishes. The second expression determines the effective equation of state. It is also useful to define the abbreviated action between the turning points,
\be
    W(\rho) \equiv \int^{\phi_2}_{\phi_1} \td \phi \frac{\partial P}{\partial \dot \phi} \ ,
\ee
since it can be used to compute all other relevant quantities:\footnote{This follows by noting that the integrand vanishes at the boundaries and that $\partial_{\rho} ( \partial_{\dot \phi} P) = \dot \phi^{-1}$.}
\be
    T_{1/2} = \partial_{\rho} W,    \qquad
    \bar P = W/T_{1/2} - \rho.
\ee
Moreover, using these relations in the time averaged continuity equation \eqref{eq:cont}, we find that $W$ is conserved in a comoving volume,
\be \label{eq:W_eom}
    \dot W + 3 H W = 0 \quad \Rightarrow \quad
    W \propto a^{-3}.
\ee
It follows that the fractional change in the energy density within a half-period is simply
\be\label{eg:Delta_rho}
    \frac{\Delta\rho}{\rho} \approx T_{1/2} \frac{\dot\rho}{\rho}  = - \frac{W}{H}.
\ee
Applying this formalism to Palatini $R^2$ models, we first note from Eq.~\eqref{eq:rho_P} that
\be\label{eq:dot_phi}
    \dot \phi^2 
    = \frac{4}{3}(\rho_0 - U) \left[\sqrt{4 - 3\frac{\rho_0 - \rho}{\rho_0 - U}} - 1 \right] \, ,
\ee
where we defined $\rho_0 \equiv (8\alpha)^{-1}$ as the initial energy density during inflation, equal to the maximum value of the potential. In the limiting cases $U \to \rho$ or $\rho \ll \rho_0$ we find $\dot \phi^2 = 2(\rho - U)$, which shows that the quartic kinetic $B$-term in \eqref{eq:rho_P} can be neglected when the motion is potential dominated or when the field has fallen off the plateau. In general, when requiring $U \leq \rho \leq \rho_0$, Eq.~\eqref{eq:dot_phi} gives
\be\label{eq:B_limit}
    \dot\phi^2  \leq \frac{1-8\alpha U}{6\alpha} \quad \Rightarrow \quad
    B \leq \frac{1}{3} \, , \quad
    \frac{2}{3} \leq c_s^2 \leq 1 \, ,
\ee
so the quartic term will also not dominate the oscillations. Numerical computations show that, in the model considered here, $B$ never saturates the upper bound \eqref{eq:B_limit}.

\begin{figure}[t]
\centering
\includegraphics[width=.8\linewidth]{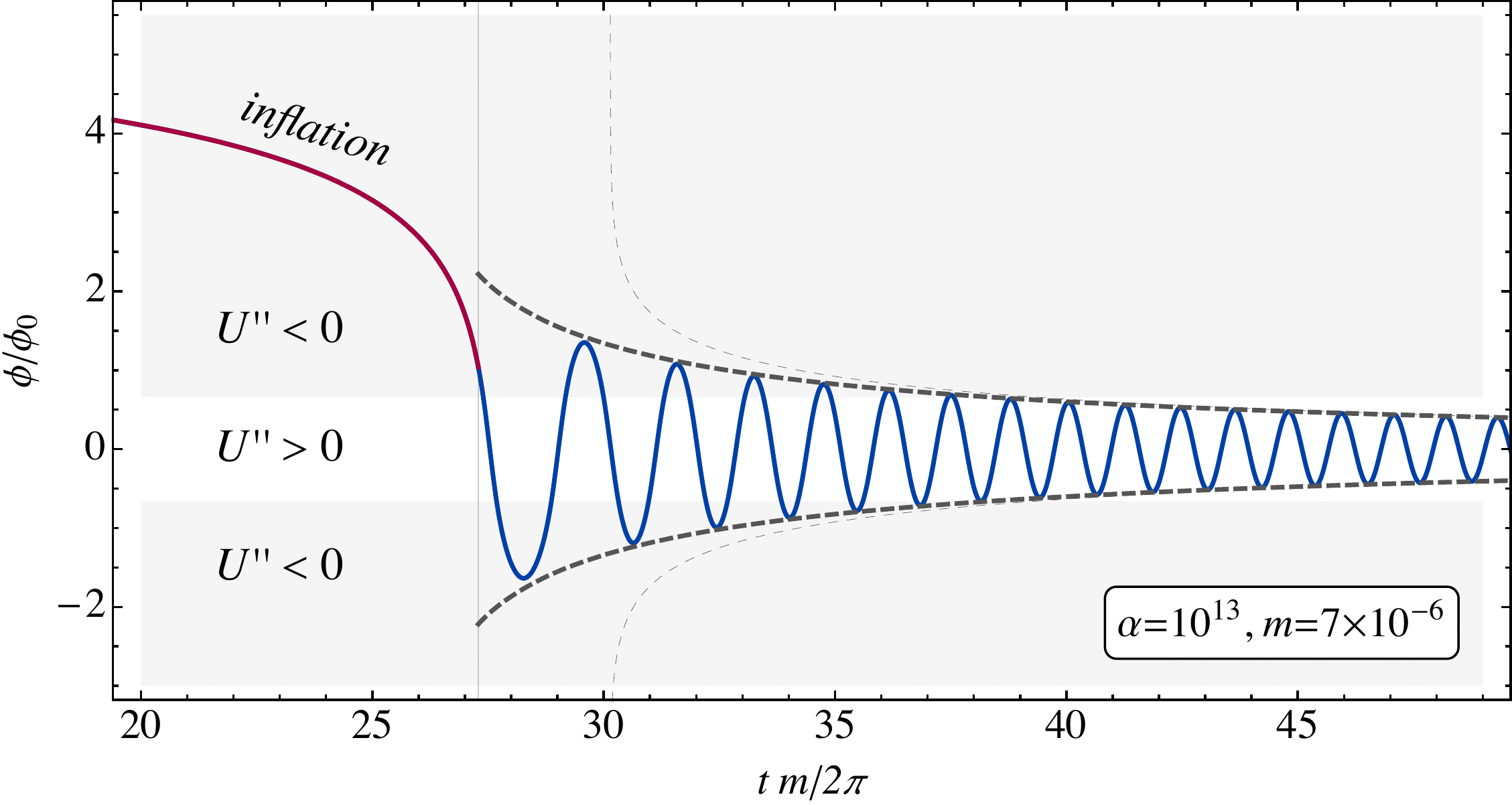}

\caption{Time evolution of the background $\phi$ (solid) just after the end of inflation with $m = 7\times 10^{-6}$ and $\alpha = 10^{13}$. The thick dashed line shows the envelope of oscillations form the analytic estimate \eqref{eq:rho_bg} while the thin dashed line shows the envelope assuming $\rho \propto a^{-3}$. The tachyonic instability is active (inactive) in the shaded (unshaded) areas.}
\label{fig:3}
\end{figure}

For our specific model with the potential \eqref{eq:pot_E}, and neglecting the subdominant $B$, we find that
\be \label{eq:W}
    W \approx \frac{\pi (1 - \sqrt{1 - \rho/\rho_0)}}{4 m \alpha} \ ,
\ee
and
\be \label{eq:T_half}
    T_{1/2} \approx \frac{\pi}{m}(1 - \rho/\rho_0)^{-1/2} \, .
\ee
We are interested in the oscillations that follow inflation, that is, we work in the regime $\rho \approx \rho_0$. In order for the formalism to apply, the energy density cannot be strongly damped by Hubble friction. Applying \eqref{eg:Delta_rho}, we see that the fractional change in the energy density within a half-period is
\be \label{eq:delta_rho}
    \frac{\Delta \rho}{\rho} 
    \approx  -\frac{W}{\sqrt{3 \rho_0}} 
    \approx -\sqrt{6}\pi \phi_0 \, .
\ee
Thus, self-consistency requires that $\phi_0 \ll 0.13$. In this case, also during the subsequent oscillations only a small fraction of the energy density is lost via Hubble damping. The tachyonic instability, $U'' < 0$ is present when $U > 1/(24 \alpha)$ or $\rho > \rho_0/3$.  As during each $n$ oscillations the amount of energy lost is roughly $2\Delta \rho$, we find that the number of oscillations in the tachyonic regime is roughly
\be\label{eq:N_tach}
    n_{\rm tach} \approx \frac{\frac{2}{3}\rho_0}{2|\Delta \rho|} \approx \frac{0.04}{\phi_0}\,,
\ee
implying that, since the mass is known \eqref{eq:m_value}, $\alpha \gtrsim 10^{13}$ is needed to make at least two oscillations before exiting the tachyonic regime. This value is 3 orders of magnitude larger than what was obtained by naive considerations in the beginning of this section. Numerical solutions (see Fig.~\ref{fig:3}) show that \eqref{eq:N_tach} slightly underestimates the actual number of oscillations in the tachyonic regime. 

The half-period of the first oscillation can be estimated by approximating the subsequent energy density as $\rho_0 - \Delta \rho$. We find that
\be
    \frac{T_{1/2}}{H^{-1}} \approx \frac{\pi \rho_0}{m\sqrt{3 \Delta \rho}} \approx 6^{-3/4}\sqrt{\pi \phi_0} \, .
\ee
Thus a small $\phi_0$ will also guarantee that a large number of oscillations will take place during one e-fold of expansion, as was assumed above.  To estimate the initial amplitude of oscillations $\phi_\mathrm{max}$, we will again use that, at the first turning point, i.e. when $\dot{\phi}=0$, the energy density is roughly $\rho_0 - \Delta \rho = U(\phi_\mathrm{max})$, which yields
\be \label{eq:phi_max}
    \phi_\mathrm{max} \approx -\frac{\phi_0}{2} \log \qty(\sqrt{\frac{3}{8}}\pi \phi_0) \, .
\ee
For $\phi_0 \ll 1$, we have $\phi_\mathrm{max} \gg \phi_0$: the field returns to the plateau after the first oscillation and also repeatedly thereafter. These estimates are in good agreement with our numerical calculations.

In the long run, neglecting fragmentation, equations \eqref{eq:W_eom} and \eqref{eq:W} imply that the energy density evolves as 
\be\label{eq:rho_bg}
    \rho \approx \rho_0 \left(\frac{a}{a_0}\right)^{-3} \left(2  - \left(\frac{a}{a_0}\right)^{-3}\right),
\ee
where $a_0$ is the scale factor at the end of inflation. The last bracketed term rapidly approaches a constant value and the Universe transitions to an effectively matter-dominated phase within less than a single $e$-fold. There $\rho \ll \rho_0$, and the half-period \eqref{eq:T_half} matches that of a harmonic oscillator. However, many oscillations can take place before this. The field repeatedly returns to the plateau where it experiences a tachyonic instability. This fragments the field condensate before the matter-dominated phase begins, as we will see in the next section. 

The envelope of the field implied by \eqref{eq:rho_bg} is shown in Fig.~\ref{fig:3} when $\alpha = 10^{13}$. Observe that even for such a large $\alpha$, the field transitions to the behaviour $\rho \propto a^{-3}$, characteristic of a massive non-interacting scalar, within the first few oscillations after inflation. For lower values of $\alpha$, this transition is even faster, reducing the effectiveness of tachyonic preheating. Therefore, considering the background evolution, it is expected that tachyonic preheating turns on around $\alpha \approx 10^{13}$. We will demonstrate this numerically in section~\ref{sec:numerical_results} (see Fig.~\ref{fig:gamma}).

\section{Preheating} 
\label{sec:particle_production}

An oscillating inflaton field generates time-dependent masses to all coupled fields and thereby induces non-perturbative particle production. This process is called preheating \cite{Traschen:1990sw, Kofman:1994rk, Shtanov:1994ce, Kofman:1997yn}, and we will study it numerically by solving the background evolution and linear perturbation equations for the inflaton.

We expand the inflaton perturbations in terms of Fourier modes $\delta \phi_k$. These follow, approximately, the mode equations
\be \label{eq:mode_eq_approx}
    \ddot{\phi}_k + 3H\dot{\phi}_k + \omega_k^2 \phi_k = 0 \ , \qquad \omega_k^2 \approx \frac{k^2}{a^2} + U'' \ .
\ee
During preheating, these perturbations grow. To quantify this growth, we compute the energy density in the perturbations, which can be estimated as
\be \label{eq:delta_rho_approx}
    \delta\rho \approx \frac{1}{2} \int^{k_\mathrm{max}} \frac{\td^3 k}{(2 \pi)^3} \left(|\delta\dot{\phi}_k|^2 + \omega_k^2 |\delta\phi_k|^2 \right) \ .
\ee
Expressions \eqref{eq:mode_eq_approx} and \eqref{eq:delta_rho_approx} are only approximations, because they neglect the effect of metric perturbations, which are coupled to the inflaton perturbations already in linear order, and the contribution from the quartic kinetic term of the inflaton. We have included these effects in our numerical computations; the more accurate (but lengthy) equations are derived in appendix 
\ref{sec:numerics}. Nevertheless, equations \eqref{eq:mode_eq_approx}, \eqref{eq:delta_rho_approx} provide a schematic understanding of the computation. In fact, they include the leading terms of the full expressions.

Traditionally, the growth of perturbations is studied in terms of the number density of particles created with respect to the adiabatic vacuum. In our case, however, the concept of particle number is ill-defined, since the effective mass $U'' < 0$ in \eqref{eq:mode_eq_approx} is tachyonic most of the time and the system never behaves adiabatically during preheating. We cannot regularize expectation values by subtracting the vacuum contribution as is usually done, since there is no adiabatic vacuum. Hence we regularize the energy density \eqref{eq:delta_rho_approx} of scalar perturbations with a UV-cutoff $k_{\rm max}$. In practice, our results are insensitive to questions about the correct vacuum or the regularization scheme, since $\delta\rho$ is dominated by growing modes below the cut-off scale $k_\mathrm{max}$. These modes grow exponentially, reaching values that far exceed the initial conditions.

After preheating ends, the particle interpretation is again valid. Assuming fast preheating and thermalization, the temperature of the particle bath is $T \sim \rho_0^{1/4} \sim (8\alpha)^{-1/4}$. This is larger than the inflaton mass $m = 7 \times 10^{-6}$~\eqref{eq:m_value} if $\alpha \lesssim 10^{20}$. We will show in section~\ref{sec:UV_cut-off} that this bound is automatically satisfied when we demand tree-level unitarity. This means that the inflaton particles are relativistic and the universe transitions into radiation domination when the inflaton fragments. Reheating is completed after the inflaton decays into standard model particles, as we discuss in section~\ref{sec:SM_interactions}.

As the inflaton's total energy density must follow the continuity equation, the fraction of the energy density in the coherent scalar will decrease due to fragmentation. This feedback can be captured at the second order of perturbation theory, but a complete description of field fragmentation requires a dedicated lattice simulation. In the following, we will limit the discussion to linear perturbations only. That is, we will establish the exponential rate of growth of the energy density in perturbations and thus the timescale of fragmentation.

\subsection{Numerical results}
\label{sec:numerical_results}

To calculate the energy density in scalar perturbations we will numerically solve the time evolution over a range of $k$-modes, for different values of the free parameter $\alpha$. Details can be found in appendix \ref{sec:numerics}. We find the spectrum of perturbations as a function of $k$, and the growth rate of the perturbation energy density for different values of $\alpha$.\footnote{In section \ref{sec:background_evolution}, we worked mostly in terms of $\phi_0$, but from this point on, we find it more convenient to use $\alpha$ and $m$. The definitions and the value $m = 7 \times 10^{-6}$ give $\phi_0 = 7 \times 10^{4} \, \alpha^{-1/2}$, so that the condition $\phi_0 \ll 1$ becomes $\alpha \gg 5 \times 10^9$.}

\begin{figure}
    \centering
    \hspace{-4mm}
    \includegraphics[width=.8\linewidth]{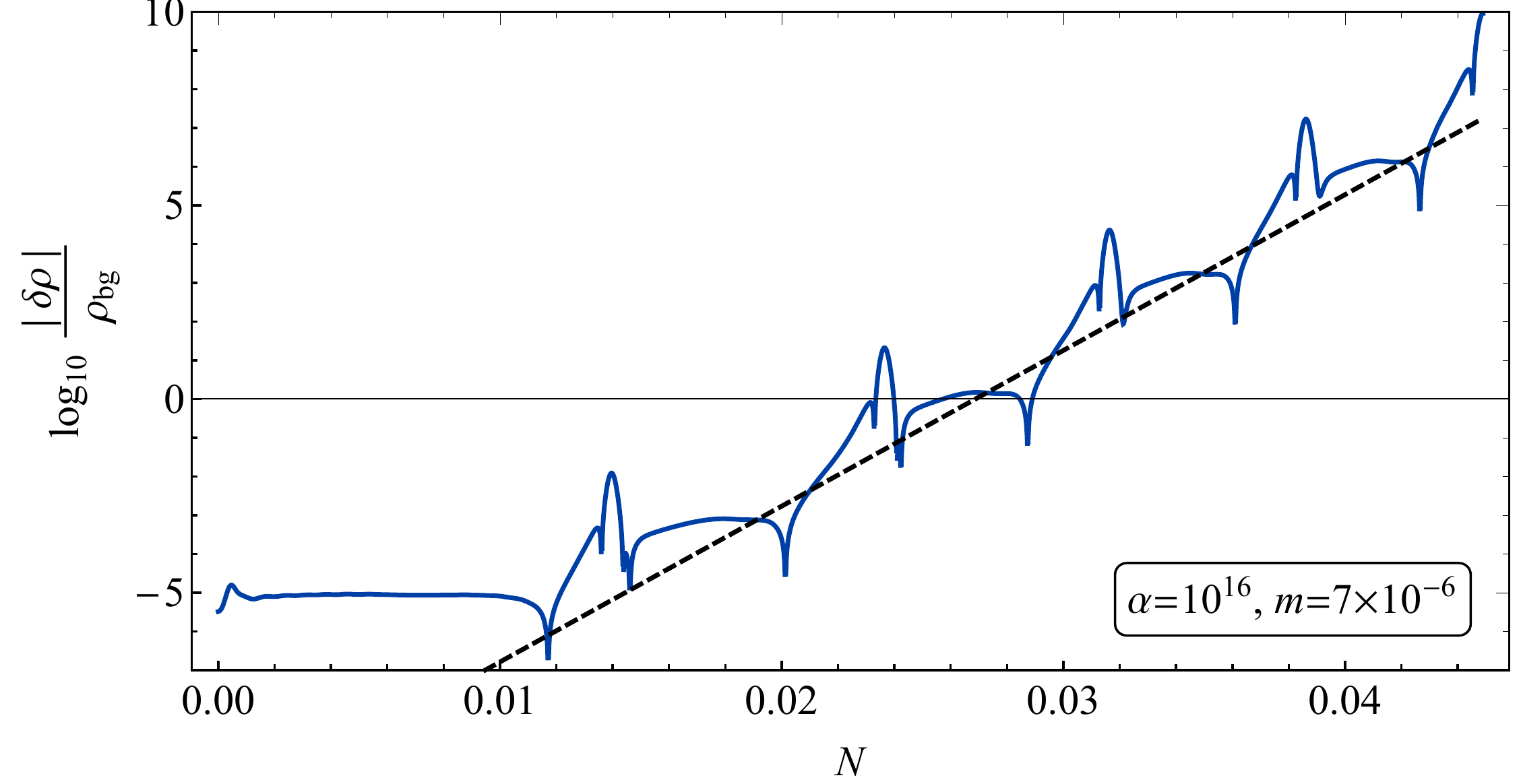}
    \hspace{4mm}
    \caption{Growth of the perturbation energy density $\delta \rho$ for $\alpha=10^{16}$, measured in units of background energy density, as a function of e-folds $N$ after the start of preheating. The strong features occur when the inflaton crosses zero and $\omega_k^2$ behaves wildly; this includes regions where $\delta \rho < 0$. The dashed line gives the average exponentially growing behaviour, denoted by $\delta \rho_\Gamma$ in the caption of Fig.~\ref{fig:spectrum}, with slope here equal to $\Gamma/(H \ln 10)$, see Eq.~\eqref{eq:Gamma}. Preheating completes in three zero-crossings, in less than one e-fold.}
    \label{fig:growth}
\end{figure}

\begin{figure}
    \centering
    \includegraphics[width=.8\linewidth]{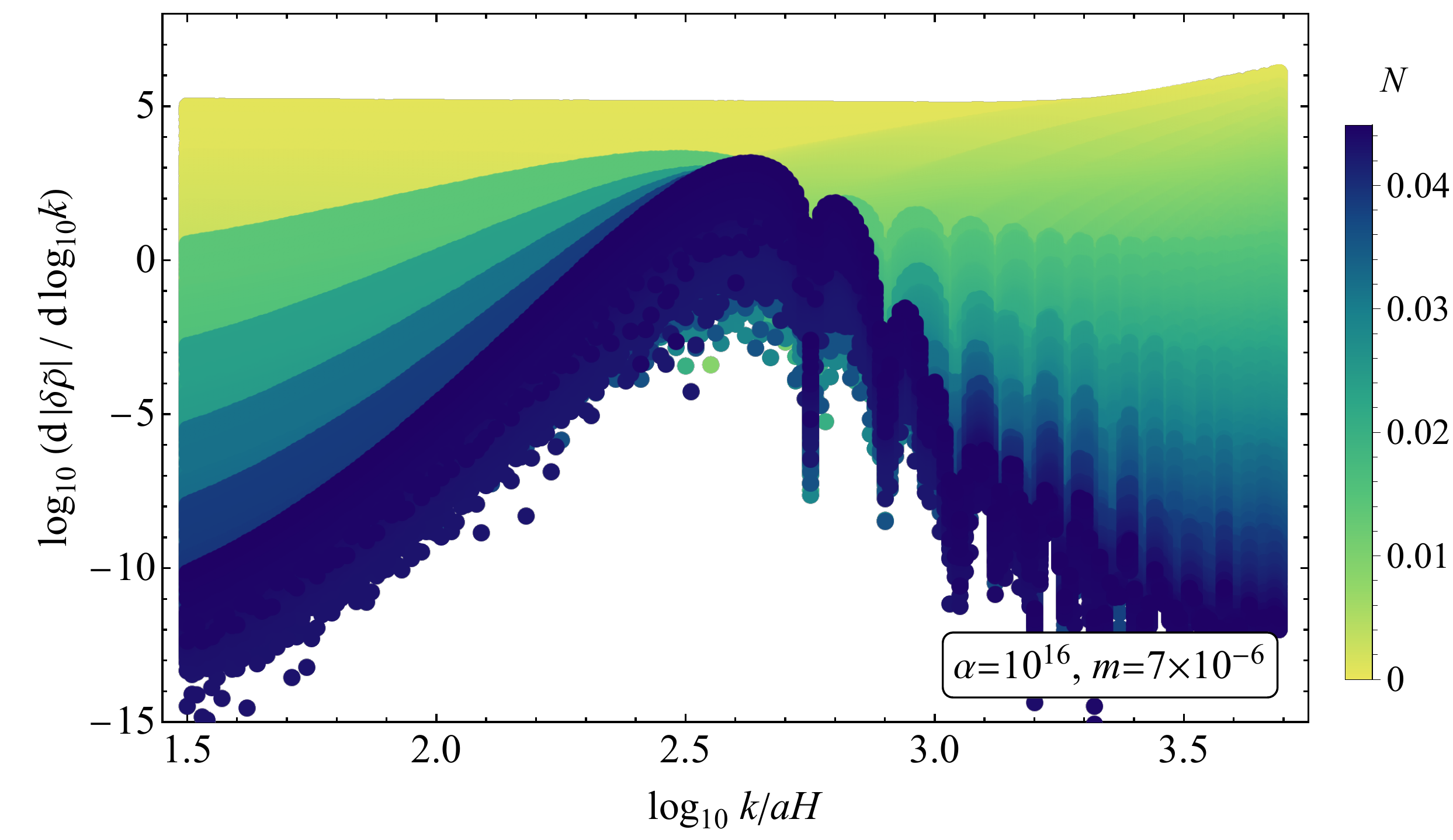}
    \caption{Rescaled energy spectrum of field perturbations at different times, with $\alpha=10^{16}$. The plotted rescaled perturbation energy density is $\delta \tilde{\rho} \equiv \delta \rho / \delta \rho_\mathrm{\Gamma}$, that is, the late-time exponential growth shown in Fig.~\ref{fig:growth} is factored out. Colours correspond to different times, measured in e-folds $N$ after the start of preheating, followed until the breakdown of perturbativity and somewhat beyond for illustrative purposes. 
    The peak of the spectrum lies deep inside the Hubble radius, around \mbox{$k \approx 400 aH$}, at the scale where tachyonic growth on the plateau is shut off. Modes with smaller $k$ grow tachyonically; modes with larger $k$ undergo non-tachyonic parametric resonance with multiple narrow resonance bands. The dominant peak modes conform to the leading exponential growth quickly; other modes grow slower, leading to a sharper and sharper peak as time goes on. The oscillations around the average behaviour seen in Fig.~\ref{fig:growth} produce the wide, overlapping `bands' of uniform colour.}
    \label{fig:spectrum}
\end{figure}

\begin{figure}
    \centering
    \hspace{-2mm}
    \includegraphics[width=.8\linewidth]{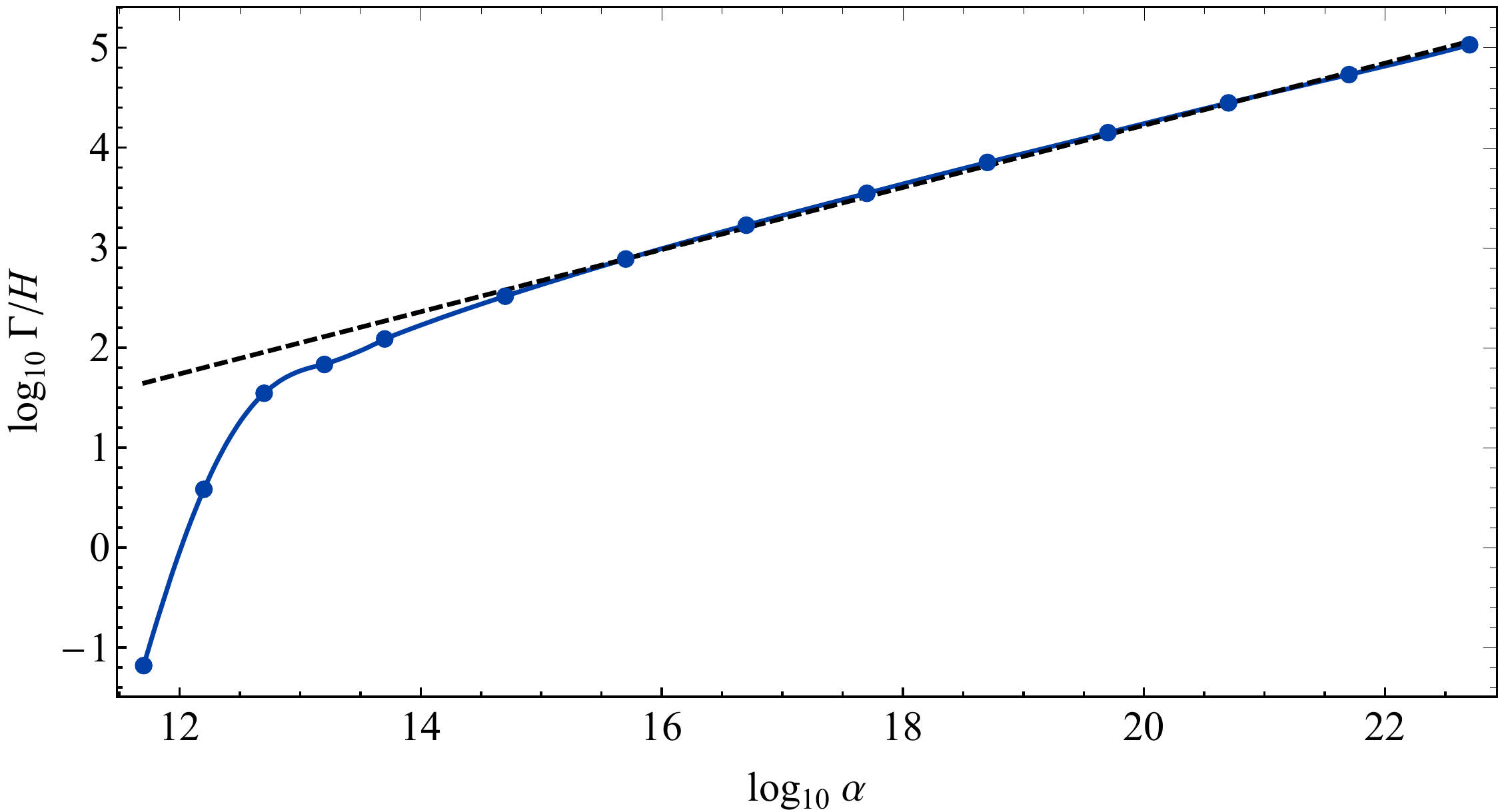}
    \hspace{2mm}
    \caption{Growth rate of perturbations as a function of $\alpha$, defined in Eq.~\eqref{eq:Gamma} and demonstrated in Fig.~\ref{fig:growth}. Its inverse $\Gamma^{-1}$ gives the timescale of preheating and background oscillations. There is a cut-off at $\alpha \sim 10^{12}$ ($\phi_0 \sim 0.1$) below which the inflaton field oscillates only near the potential's quadratic minimum with no inflaton particle production. Above this, $\Gamma/H \approx 0.011 \times \alpha^{0.31}$, indicated by the dashed line. Our numerical calculations break down above $\alpha \sim 10^{23}$ due to insufficient numerical accuracy, but these values also run into problems with unitarity, see Eq.~\eqref{eq:alpha_condition}.}
    \label{fig:gamma}
\end{figure}

As already mentioned, the inflaton field probes the plateau repeatedly during its oscillations during preheating. There the inflaton mass $U''$ is negative, leading to tachyonic, exponential growth of the perturbations, as previously studied in \cite{Rubio:2019ypq, Karam:2020rpa}. All modes for which $\omega_k^2$ from \eqref{eq:mode_eq_approx} is negative when the background field is at its oscillation amplitude $\phi_\mathrm{max}$ undergo tachyonic growth: they spend most of the time in the instability regime. For large enough $k$, this is no longer true. The fastest-growing peak modes occur at the largest $k$-values which are still tachyonic, that is,
\be \label{eq:k_peak_approx_1}
    \frac{k_\mathrm{peak}}{a} \approx \sqrt{-U''(\phi_\mathrm{max})} \, .
\ee
Using the analytical approximation \eqref{eq:phi_max} for $\phi_\mathrm{max}$  gives the estimate
\be \label{eq:k_peak_approx_2}
     \frac{k_\mathrm{peak}}{aH} \approx 2\times6^{3/4}\sqrt{m\pi} \alpha^{1/4} \approx 0.036 \times \alpha^{1/4} \, .
\ee
By performing a numerical fit we find that, for large $\alpha$, the peak frequency scales as
\be \label{eq:k_peak_numerical}
    \frac{k_\mathrm{peak}}{aH} \approx 0.030 \times \alpha^{0.26} \, ,
\ee
which is in good agreement with the analytic estimate. In Fig.~\ref{fig:spectrum} we see that the spectrum of energy density falls off sharply on both sides of the peak. The momentum cutoff $k_\mathrm{max}$ from Eq.~\eqref{eq:mode_eq_approx} is chosen so that $k_\mathrm{max} \gg k_\mathrm{peak}$ and the full peak with all out-of-vacuum modes is accounted for in the integrated perturbation energy density.

Radiation formed at the peak scale will be redshifted and, at present, will carry the frequency
\begin{equation}
    f = \frac{1}{2\pi}\frac{k_\mathrm{peak}}{ a_\mathrm{now}} \approx 0.002 \, T_\mathrm{0} \approx 1 \, \text{GHz} \, ,
\end{equation}
where $a_\mathrm{now}$ is the scale factor today, $T_\mathrm{0} = 2.7$~K is the current temperature of radiation, and we used $\rho \approx T^4$ after preheating and $a \propto 1/T$. All $\alpha$-dependence cancels out. The violent preheating process may produce strong gravitational waves at this scale. The frequency is beyond the reach of current or near-future gravitational laser-interferometric GW detectors which are sensitive to frequencies up to 10kHz~\cite{Somiya:2011np,TheVirgo:2014hva,Martynov:2016fzi}. Currently, the most sensitive probe of such primordial GW signals is the number of effective relativistic degrees of freedom $N_{\rm eff}$ during Big Bang nucleosynthesis~\cite{Pagano:2015hma,Ejlli:2019bqj,Domcke:2020yzq}. However, a GW background at 3-30GHz frequency range can also leave an imprint on the CMB Rayleigh-Jeans tail and may thus be probed via future 21cm physics experiments~\cite{Domcke:2020yzq,Ringwald:2020ist}.

Integrating over the spectrum gives the total perturbation energy density, which also grows exponentially, see Fig.~\ref{fig:growth}, and is dominated by the peak mode. For large $\alpha$, the process is efficient: the inflaton condensate fragments within less than a single e-fold of expansion and takes only a few background oscillations. We estimate the timescale of fragmentation as the rate of the growth of perturbation energy density:
\be\label{eq:Gamma}
    \Gamma \equiv \frac{d \ln \delta \rho}{d t} \, .
\ee
This is essentially constant around the time when the perturbation energy density crosses the background energy density. Note that $\Gamma^{-1}$ gives the timescale of preheating, and thus the timescale for the transition from inflation to radiation domination.

In Fig.~\ref{fig:gamma} we plot $\Gamma/H$ as a function of $\alpha$. It follows a broken power law. Above $\alpha \sim 10^{12}$, 
\be \label{eq:GammaH}
    \frac{\Gamma}{H} \approx 0.011 \times \alpha^{0.31} \, .
\ee
Part of the $\alpha$-scaling arises from $H \propto \alpha^{-1/2}$ at the end of inflation, while the rest is due to the effect of $\alpha$ on the evolution of the oscillations. When $\alpha \lesssim 2 \times 10^{12}$, the background field no longer returns to the region at which $U''<0$, and the tachyonic production shuts off. There is no fragmentation of the inflaton condensate, since the inflaton mass is constant near the quadratic potential minimum, preventing parametric resonance. Reheating must then proceed through other channels, such as non-perturbative production of other fields or the perturbative decay of the inflaton. These details are highly model-dependent. Effective preheating through inflaton fragmentation then requires $\alpha \gtrsim 10^{13}$, in agreement with our previous estimates from section~\ref{sec:oscillation}. Note that this gives $r \lesssim 4 \times 10^{-7}$, which is too small to be probed in the near future. 

We repeated the computation with the simplified expressions \eqref{eq:mode_eq_approx} and \eqref{eq:delta_rho_approx}, dropping the quartic kinetic term also from the background equations, and compared the obtained $\Gamma$-values to the accurate results. For $\alpha \gtrsim 10^{13}$, the difference is less than $10\%$, showing that both the metric perturbations and the quartic kinetic term play subleading roles during preheating.

In summary, we have shown that the inflaton field $\phi$ quickly fragments into inflaton particles and the universe transitions into radiation domination. Reheating can then be completed through the decay of the inflaton into Standard Model particles. We will discuss this in section \ref{sec:SM_interactions}, after commenting on the UV cut-off scale of the model.

\subsection{UV cut-off} \label{sec:UV_cut-off}

The Lagrangian \eqref{action6} for the scalar field is non-renormalizable due to the quartic kinetic term and the non-polynomial potential. The non-renormalizable contributions may cause problems such as violation of tree-level unitarity in high-energy scattering. The quartic kinetic term $~\alpha (\partial\phi)^4$ implies that, for the perturbations, the problems arise at energy scale $\alpha^{-1/4}$, while the relevant energy scale related to the potential $U$ is $\phi_0$. To describe preheating consistently, the energy scale of the produced particles should remain below both of those cut-off scales.

The energy scale of the produced inflaton particles is, by Eq.~\eqref{eq:k_peak_numerical}, $k_\mathrm{peak}/a \approx 0.007 \alpha^{-1/4}$. After thermalization, the typical energy scale of a particle is $\rho_0^{1/4} \approx 0.6 \alpha^{-1/4}$, assuming thermalization is fast and the energy density does not dilute considerably during it. We see that both of these scales are below the cut-off scale arising from the quartic kinetic term (if only barely for $\rho_0^{1/4}$). To push them below the $U$-cut-off as well, we demand
\begin{equation} \label{eq:alpha_condition}
    \alpha^{-1/4} \lesssim \phi_0 \, \qquad \iff \qquad \alpha \lesssim m^{-4} \sim 10^{20} \, .
\end{equation}
As long as this limit is obeyed, our preheating treatment is consistent.
We remark that the upper bound on $\alpha$ implies that $r > 4 \times 10^{-14}$, and is thus in conflict with proposals relying on Palatini $R^2$ inflation~\cite{Tenkanen:2019wsd, Tenkanen:2020cvw} for accommodating the ``Trans-Planckian Censorship" conjecture which suggests $r \lesssim 10^{-30}$~\cite{Bedroya:2019tba}\footnote{See, however, ref.~\cite{Mizuno:2019bxy}.}.

\subsection{SM interactions} \label{sec:SM_interactions}

In Palatini $R^2$ gravity, the conformal transformation~\eqref{eq:conf} will couple the inflaton to every non-conformally coupled field in the Einstein frame. In the SM, the only non-conformal field is the Higgs field. It is thus necessary to study its couplings.

Since the scale of inflation is much higher than the electroweak scale, we neglect the Higgs mass term and take a Higgs potential of the form
\be 
    V(h) =  \frac{\lambda_h}{4} h^4 + \frac{\lambda_{h\varphi}}{4} h^2 \varphi^2 \,,
\ee
where $\lambda_{h}$ is the quartic coupling of the Higgs and $\lambda_{h\varphi}$ is a portal coupling between the Higgs and the inflaton. After transforming into the Einstein frame, the action for the inflaton-Higgs sector is
\be \label{eq:higgs_action}
    S
    =   \int\dd^4 x \sqrt{-g} \left\{ \frac{1}{2} R - \frac{1}{2} (1 - 8\alpha U) \qty[\left( \partial \varphi \right)^2 + \qty(\partial h)^2] + \frac{\alpha}{2} (1 - 8\alpha U) \qty[\left( \partial \varphi \right)^2 + \qty(\partial h)^2]^2  - U \right\} \ ,
\ee
with the potential
\be \label{eq:UtotHiggs3}
U = \frac{\frac{1}{2} m^2 \varphi^2  + \frac{\lambda_h}{4} h^4 + \frac{\lambda_{h\varphi}}{4} h^2 \varphi^2}{1 + 8 \alpha \left( \frac{1}{2} m^2 \varphi^2  + \frac{\lambda_h}{4} h^4 + \frac{\lambda_{h\varphi}}{4} h^2 \varphi^2  \right)} \,.
\ee
This is equivalent to the Jordan frame action \eqref{action5} but with the Higgs field and a potential added on an equal standing with the inflaton.
The inflaton's kinetic term can be made canonical with the same field redefinition as earlier, using $h=0$ in the defining equation \eqref{fieldredef}. For the Higgs, the full quadratic kinetic term reads
\begin{equation} \label{eq:higgs_kinetic}
    -\frac{1}{2} \qty(\partial h)^2 \times \qty( 1-8\alpha U_0 + 2\alpha \dot{\phi}^2) \, ,
\end{equation}
where $U_0 \equiv \frac{1}{8\alpha}\tanh^2 \qty(\phi/\phi_0)$, the background potential without $h$ or inflaton perturbations; similarly, $\dot{\phi}$ is the background field velocity here and in the following paragraphs. We can make the Higgs field (almost) canonical by the field redefinition
\begin{equation} \label{eq:s}
    s \equiv h\sqrt{1-8\alpha U_0 + 2\alpha \dot{\phi}^2} \, .
\end{equation}
The resulting kinetic term is not fully canonical, since $\phi$ depends on time, which results in extra terms of the form $\dot{\phi}\dot{s} s$ when taking time derivatives. However, this kinetic term is sufficient for order-of-magnitude estimates. In particular, we can derive mode equations for $s$ which take the standard form:
\begin{equation} \label{eq:s_mode_eq}
    \ddot{s}_k + 3H\dot{s}_k + \qty(\frac{k^2}{a^2} + m^2_\mathrm{eff,s}) s_k = 0 \, .
\end{equation}
The complete expression for the effective mass squared $m^2_\mathrm{eff,s}$ is messy, but it contains two types of terms: those arising from the potential,
\begin{equation} \label{eq:s_mass_1}
    m_{U,s}^2 = \frac{\lambda_{h\varphi} U_0 (1-8\alpha U_0)}{m^2\qty(1-8\alpha U_0 + 2\alpha \dot{\phi}^2)} > 0 \, ,
\end{equation}
and terms arising from the kinetic sector, $m_\mathrm{kin,s}^2 \sim \dot{\phi}^2$. All terms containing $\dot{\phi}$ are slow-roll suppressed during inflation. To ensure the stability of the Higgs field there, we demand
\begin{equation} \label{eq:lambda_hphi_condition_1}
    m_{U,s}^2 \approx \frac{\lambda_{h\varphi}}{8\alpha m^2} > H^2 \approx \frac{1}{24\alpha} \,  \qquad \Rightarrow \qquad \lambda_{h\varphi} > \frac{m^2}{3} \sim 10^{-11} \, .
\end{equation}
After inflation, the kinetic mass terms are no longer suppressed, but they all have the same form,
\begin{equation}
    m_\mathrm{kin,s}^2 \sim \qty(\frac{\dot f}{f})^2 \, ,
    \qquad
    \dot{f} \sim \frac{\dot{\phi}}{\phi_0}f \, , \qquad f \sim 1
    \qquad
    \Rightarrow
    \qquad
    m_\mathrm{kin,s}^2 \sim \frac{\dot{\phi}^2}{\phi_0^2} \sim \frac{m^2 \alpha}{\alpha} \sim m^2 \, .
\end{equation}
Here $f$ refers to a class of hyperbolic and exponential functions, such as $8\alpha U_0 = \tanh^2 (\phi/\phi_0)$ and its derivatives, which follow the given properties during the oscillations. We also used $\dot{\phi}^2 \sim \rho_0\sim  1/\alpha$: the kinetic energy of the ocillations is the same order as the potential energy. The $m_\mathrm{kin,s}^2$ terms can be negative and cause a strong tachyonic instability to the Higgs (see \cite{Karam:2020rpa} for a detailed study of a similar case). However, this tachyonicity can be shut off by making the positive mass term $m_{U,s}^2$ large enough:
\begin{equation} \label{eq:lambda_hphi_condition_2}
    m_{U,s}^2 \sim \frac{\lambda_{h\varphi}}{\alpha m^2} > |m_\mathrm{kin,s}^2|  \sim m^2 \qquad \Rightarrow \qquad
    \lambda_{h\varphi} > m^4 \alpha \sim \alpha \times 10^{-20} \, .
\end{equation}
This condition ensures that no tachyonic Higgs production takes place. Higgses can still be produced through parametric resonance, but this is likely to be a subleading channel compared to the fast tachyonic production of the inflaton. Note that \eqref{eq:lambda_hphi_condition_2} can be satisfied for a perturbative value $\lambda_{h\varphi} < 1$ as long as $\alpha$ satisfies the condition \eqref{eq:alpha_condition}.

For the interactions, all new non-renormalizable terms involving the Higgs boson come from expanding $8\alpha U$, see \eqref{eq:higgs_action} and \eqref{eq:UtotHiggs3}. These consist of renormalizable operators multiplied by powers of $\alpha m^2\varphi^2$, implying the cut-off scale $(\alpha m^2)^{-1/2} \sim \phi_0$, and powers of $\alpha \lambda_h h^4$ and $\alpha \lambda_{h\varphi} h^2 \varphi^2$, with the implied cut-offs $(\alpha\lambda)^{-1/4} > \alpha^{-1/4}$ for perturbative $\lambda$. These order-of-magnitude estimates do not change when switching to the Einstein frame fields $\phi$ and $s$, since both the Jordan and the Einstein frame fields are of the same order during preheating. These cut-offs are larger than those discussed in section \ref{sec:UV_cut-off}, so the treatment is still consistent as long as the $\alpha$-bound \eqref{eq:alpha_condition} is obeyed.

The emerging picture is then the following: with the bounds \eqref{eq:lambda_hphi_condition_1}, \eqref{eq:lambda_hphi_condition_2}, Higgses are not produced during inflation, and the violent tachyonic inflaton production still dominates preheating. After preheating, the inflaton can decay into the SM through the Higgs portal coupling and complete reheating. We have no upper bound for $\lambda_{h\varphi}$ (beyond demanding perturbativity), so this decay can be efficient.

\section{Conclusions}
\label{sec:conclusions}

We studied preheating in the Palatini formulation of general relativity, focusing on the model with a quadratic inflaton potential and an $\alpha R^2$ term in the action. In the Einstein frame, this is a plateau model compatible with the CMB observations. In this model, however, the inflaton's kinetic term has a quartic component, making the analysis more involved than in standard plateau inflation. 

We showed that for sufficiently large values $\alpha \gtrsim 10^{13}$ the inflaton returns to the plateau repeatedly during preheating, giving rise to a tachyonic instability similar to that of previous Palatini studies \cite{Rubio:2019ypq, Karam:2020rpa}. This leads to a fast fragmentation of the inflaton field and the onset of radiation domination in less than a single e-fold of spatial expansion. We obtained the spectrum of the perturbations numerically for different values of $\alpha$. When $\alpha \gtrsim 10^{13}$, the spectrum peaks at the comoving wavenumber 
\be
    k_\mathrm{peak}/aH \approx 0.030 \times \alpha^{0.26}\nonumber \,,
\ee
deep within the Hubble radius, and the energy density of the fragmented field grows exponentially with the rate
\be
    \Gamma/H \approx 0.011 \times \alpha^{0.31}\nonumber \,,
\ee
which is considerably larger than the rate of expansion. Below $\alpha \approx 10^{13}$ the growth rate of the perturbations begins to drop fast. In the range $10^{12} \lesssim \alpha \lesssim 10^{13}$ preheating is still possible, but with a greatly reduced rate. When possible interactions with the SM Higgs are included, the tachyonic inflaton production still tends to dominate preheating.

Our numerical studies show that neglecting the Einstein frame quartic kinetic term does not significantly affect the evolution of the smooth background (both during inflation or the later coherent oscillations) or the dynamics of preheating. In particular, the effect on $\Gamma$ is less than 10\%. This complements the earlier results that the quartic kinetic term is subleading during slow-roll inflation \cite{Enckell:2018hmo, Tenkanen:2020cvw}. While the study of preheating dynamics was mostly numerical, we provided a more complete analytic picture of the evolution of the oscillating background field. In particular, we prove analytically that the quartic kinetic term must always be smaller than the quadratic one.

A general consequence of a large $\alpha$ in Palatini $R^2$ models is a strong suppression of the tensor-to-scalar $r$. We show that unitarity considerations together with an effective preheating suggest the range $10^{13} \lesssim \alpha \lesssim 10^{20}$, implying a tensor-to-scalar ratio $4\times 10^{-14} \lesssim r \lesssim 4\times10^{-7}$, too low to be detectable in near-future experiments.

Our study was limited to linear perturbations, omitting the interactions between the perturbations and the background. The inclusion of such effects requires a lattice simulation. Previous lattice studies of similar models \cite{Lozanov:2016hid, Lozanov:2017hjm, Krajewski:2018moi, Lozanov:2019ylm}, albeit without the quartic kinetic terms, support our results of a fast, efficient preheating process.

\acknowledgments
This work was supported by the Estonian Research Council grants PRG803, PRG1055, MOBJD381 and MOBTT5 and by the EU through the European Regional Development Fund CoE program TK133 ``The Dark Side of the Universe."

\appendix

\section{Quartic kinetic term during inflation}
\label{sec:quartic_kinetic_term}

In order to consistently neglect the contribution of the quartic kinetic term in \eqref{action6} during inflation, the function $B$ \eqref{eq:B_cs} must be small. To estimate the size of $B$ during slow-roll, consider the field equations in  terms of $e$-folds $N \equiv \ln a$. Since\footnote{In this appendix, prime denotes a derivative with respect to $N$, except in $U'$, which still denotes $\frac{\mathrm{d}U}{\mathrm{d}\phi}$.} $\dot\phi = H\phi'$ and $B \propto H^2$, the Friedmann equation gives a quadratic equation for $H^2$ that can be used to recast the dynamics in terms of $\phi$, $\phi'$ only. After some algebra we see that the (exact) evolution of the homogeneous background is determined by
\bea
    \phi'' + \phi' \left( 3 c_s^2 - \eps_1\right) + \frac{1 + 3B^2}{1 + 3B} \frac{U'}{H^2} = 0 \, ,
\eea
with
\bea\label{eq:B_N}
    B 
&   = \frac{8\alpha U}{1 - 8 \alpha U} \frac{\phi'^2}{6-\phi'^{2}}\left(1 + \sqrt{1 - \frac{24 \alpha U}{1 - 8\alpha U} \frac{\phi'^{4}}{(6-\phi'^{2})^2}} \right)^{-1} \,.
\eea
The Hubble parameter $H$ and the first slow-roll parameter $\eps_1$ read
\be
    H^2 
   = \frac{U}{3 - \phi'^2/2 - 3B \phi'^2/4}, \qquad
    \eps_1 
   \equiv -\frac{H'}{H} = \frac{1+B}{2} \phi'^2.
\ee
We remark that in case the field redefinition \eqref{fieldredef} does not admit an analytic form, then, in practical computations, the dynamics is more conveniently described by first order equations using the phase space variables $\varphi$ and $y \equiv \phi'$ and the additional equation $y = \sqrt{ \frac{1 + 8\alpha U}{1+A}} \varphi'$.

Returning to the model with a quadratic potential \eqref{eq:pot_J}, the slow-roll approximation gives
\be\label{eq:N_SR_2}
    N 
    = \int^{\phi}_{\phi_e} \frac{\dd \phi}{\sqrt{2 \epsilon_U}}
    = \frac{\phi_0^2}{4}\sinh^2{\frac{\phi}{\phi_0}} \, ,
\ee
where $\phi_e$ is the field value at the end of inflation. As a rough approximation we use $\phi_e = 0$ as the corresponding error in $N$ is less than an e-fold. Plugging $\phi(N)$ into \eqref{eq:B_N} and evaluating the Pad\'{e} approximant of the order [1/1] for large $N$ gives
\be
    B \approx \frac{1}{12 N + 3\phi_0^2 -3/2}.
\ee
Thus, the quartic kinetic term can be consistently ignored during inflation when $N \gg 1$.

\section{Solving for cosmological perturbations}
\label{sec:numerics}

Scalar perturbations of a spatially flat FRW metric in the longitudinal gauge are captured by~\cite{Mukhanov:1990me}
\be
    \td s^2 = \left(1 + 2\Phi \right) \td  t^2 - a^2 \left(1 - 2\Psi\right)\delta_{ij} \td  x^i \td  x^j \,,
\ee
where $\Phi$ and $\Psi$ are the Bardeen potentials and the perturbations of the scalar field are $\delta\phi \equiv \phi - \bar{\phi}$, where $\bar{\phi}$ denotes the background field. In terms of the Sasaki-Mukhanov variable $v \equiv \left( \Phi + H \delta\phi / \dot\phi \right)\, z $ and the conformal time $\td\tau=\td t/a$, the dynamics of scalar perturbations is described by the action~\cite{Garriga:1999vw}
\be
    S = \frac{1}{2} \int  \left( v_{,\tau}^2 - c_s^2 (\nabla v)^2 + \frac{z_{,\tau\tau}}{z} v^2 \right) \td\tau \td^3\mathbf{x} \ ,
\ee
where
\be
    z \equiv \frac{a\dot\phi}{c_s H} \sqrt{1 + B} = a \phi' \sqrt{1 + 3B} \ .
\ee
The momentum eigenmodes obey the Sasaki-Mukhanov equation
\be \label{eq:MS}
    v_{,\tau\tau} + \omega_k^2 v = 0, \qquad
    \omega_k^2 \equiv c_s^2 k^2 - \frac{z_{,\tau\tau}}{z} \ ,
\ee
where the expression for $z_{,\tau\tau}/z$ can be expressed as
\be 
    z_{,\tau\tau}/z 
    = a^2 H^2 \left( 2 - \epsilon_1 + \frac{3}{2} \epsilon_2 - 3 s - \frac{1}{2} \epsilon_1 \epsilon_2 + \epsilon_1 s - \epsilon_2 s + \frac{1}{4} \epsilon^2_2 + s^2 + \frac{1}{2} \epsilon_2 \epsilon_3 - \frac{\dot{s}}{H}  \right) \ ,
\ee
where
\begin{equation} \label{eq:epsilons}
    \epsilon_1 \equiv - \frac{\dot H}{H^2} = - \frac{\dd \ln H}{\dd N} \ , \quad
    \epsilon_{n+1} \equiv \frac{\dd \ln \epsilon_n}{\dd N} = \frac{\dot{\epsilon}_n}{H \epsilon_n} \ , \quad
    s \equiv \frac{\dot{c}_s}{H c_s} \ .
\end{equation}
Sufficiently far past in the inflationary epoch any mode is within the horizon, $k \gg a H $. In this limit, $\omega_k^2 \approx c^2_s k^2$ and, assuming a slowly varying speed of sound $c_s$, the Mukhanov-Sasaki equation~\eqref{eq:MS} is solved by $v(\tau) = e^{-i c_s k \tau}/\sqrt{2 c_s k}\,$, defining the so-called Bunch-Davies vacuum. These equations provide both the primordial power spectrum of curvature perturbation generated during the inflationary epoch,
\be
    \mathcal{P}_{\mathcal{R}} (k) \equiv \frac{k^3}{2\pi^2} \Big\vert \frac{v}{z}  \Big\vert^2 \,,
\ee
as well as the preheating dynamics when the perturbations are in the linear regime.

We compute the energy density of field perturbations in the FRW background, defined as
\be
    \delta\rho = T_{00}|_\text{pert} = \frac{-2}{\sqrt{-g}}\frac{\delta S}{\delta g^{00}}\Bigg|_\text{pert} \ ,
\ee
expanded to second order in field perturbations and with the homogeneous background contribution removed. We neglect metric perturbations when calculating $\delta \rho$. Using the expectation value of the corresponding quantum operator in the Bunch-Davies vacuum (see \cite{Birrell:1982ix}), the first-order terms vanish and the second-order terms can be written in terms of the Fourier mode functions of the field, $\phi_k$. We have that
\bea \label{eq:rho_full}
    \expval{\delta \rho} &= \int_{k_\mathrm{min}}^{k_\mathrm{max}} \frac{\dd \ln k \ k^3}{2\pi^2} \delta \rho_k,
\eea
where
\bea
   \delta \rho_k 
&   \equiv \frac{1}{2}\qty(1 + 9B) |\dot{\phi}_k|{}^2 +12B\tilde B \Re \phi_k \dot{\phi}_k^* \\
&   + \frac{1}{2}\qty[\frac{k^2}{a^2}\qty(1 - B) 
    + U'' (1 + 3B^2) + 24 B \tilde B^2] |\phi_k|{}^2 \, 
\eea
and we defined $\tilde B \equiv 2 \alpha \dot{\phi} U'/(1 - 8\alpha U)$. The integral is regulated by cut-offs which remove the adiabatic high-$k$ modes ($k>k_\mathrm{max}$) and the frozen super-Hubble modes ($k < k_\mathrm{min}$) which are not excited during preheating. To study how the inflaton condensate fragments into the perturbations, the resulting quantity can then be compared to the background energy density.

Expression \eqref{eq:rho_full} does not include metric perturbations and is therefore not a fully rigorous description of the energy density of the perturbation. However, it should be a good measure of the fragmentation of the inflaton condensate when the field perturbation is computed in the spatially flat gauge, where metric perturbations are essentially minimized. Our numerical results show that \eqref{eq:rho_full} is almost identical to the `naive' perturbation energy density introduced in Eq.~\eqref{eq:delta_rho_approx}. This equivalence holds up to fluctuating order-one corrections which are not important when determining the exponential growth rate $\Gamma$ in \eqref{eq:Gamma}.

To compute \eqref{eq:rho_full}, we numerically solve the mode equation for $\phi_k$ in the spatially flat gauge in which the curvature perturbation is zero, so that the Sasaki-Mukhanov variable is $v=a \, \delta\phi \sqrt{1+3B}$. We derive the mode equation 
\be
    \ddot{\phi}_k + f\dot{\phi}_k + \omega_k^2 \phi_k = 0
\ee
from the Sasaki-Mukhanov equation \eqref{eq:MS}. The coefficients read
\bea \label{eq:mode_eq_full}
    f &= 3H\frac{1+3B^2}{(1+3B)^2} - \frac{6(1+B)(1-3B)\tilde B}{(1+3B)^2},
    \\
    \omega_k^2 &= \frac{k^2}{a^2}c_s^2 + \frac{1+3B^2}{1+3B} U''
    -\frac{\dot{\phi}^4}{2 H^2} (1 +B)^2
    +\frac{2 \dot{\phi} U'}{H}\frac{(1 + B)(1 + 5 B)}{(1 + 3 B)^2}
   \\
&   + 3\dot{\phi}^2\frac{(1 + 6B)(1  + B)^2}{(1 + 3B)^2}
    - 24 H \frac{B \tilde B}{(1 + 3 B)^2}
    - \frac{12 (1 + B)(1 - 3B)\tilde B^2}{(1 + 3B)^2}\, .
\eea
Numerical tests confirm that the leading contribution to $\omega_k^2$ coincides with the `naive' expression $k^2/a^2 + U''$ in Eq.~\eqref{eq:mode_eq_approx}. The other terms, arising from the non-canonical kinetic terms and the coupled metric perturbations, are sub-leading and can be neglected to a good accuracy when calculating the growth rate $\Gamma$ and the shape of the spectrum. Similarly, the friction term with $f$ is negligible. Thus, the non-canonical kinetic terms are unimportant during preheating, just as they were shown to be unimportant during slow-roll in \cite{Enckell:2018hmo, Tenkanen:2020cvw}. The irrelevance of the metric perturbations was also noted in the context of Higgs inflation in \cite{Rubio:2019ypq}, suggesting that neglecting the metric perturbations when computing the energy density \eqref{eq:rho_full} is justified.

To solve for the time evolution of the energy density of the perturbations, we solve the mode equation \eqref{eq:mode_eq_full} with the Bunch-Davies initial conditions during inflation for a range of modes between $k_\mathrm{min}$ and $k_\mathrm{max}$ and use these to compute the integral \eqref{eq:rho_full}. Its time evolution is followed until it becomes comparable to the background energy density. From this time evolution the growth rate $\Gamma$ \eqref{eq:Gamma} is extracted via a numerical fit, see Fig.~\ref{fig:growth}. This process is repeated for different values of the $\alpha$-parameter to produce Fig.~\ref{fig:gamma}.

\bibliography{References}

\end{document}